# ESA Science Programme Missions: Contributions and Exploitation – Payload Provision

John C. Zarnecki · Arvind N. Parmar

**Abstract** We have collected data pertaining to the Principal Investigators (PIs), and co-PIs (where appropriate) for all ESA-led Science Directorate missions since the first such launch, namely of COS-B in 1975. For a total of 28 missions (including 4 in preparation awaiting launch), 437 individuals have been recorded along with their institution, location, "academic age" and gender. We have correlated the number of PIs by country with the financial contribution of those countries to the ESA Science programme. We have also investigated issues associated with age and gender of the PIs. As a result of these analyses, we make suggestions for actions which ESA and its Member States may wish to consider with the aim of encouraging equity and diversity while still placing scientific excellence as the overarching goal.

## 1 Introduction

European Space Agency (ESA)'s Science Directorate missions have, over the last five decades, followed a variety of formats. However, two distinct patterns can be discerned from the large number of missions flown. At one end of the spectrum there are missions which incorporate a collection of separate instruments, or payloads, generally functioning independently of each other, at least from a technical perspective, although very often the scientific aims and objectives are interrelated, sometimes very closely. Very occasionally there may be a closer

J. Zarnecki
The Open University
Milton Keynes
UK
E-mail: j.c.zarnecki@open.ac.uk

A. Parmar
Former Head of the Science Support Office
Directorate of Science, ESA, ESTEC
The Netherlands
*Present address:*
Department of Space and Climate Physics
MSSL/UCL
Dorking
UK

relationship between two such instruments – an example of such a dependency occurred on the Huygens Probe [1] where the Aerosol Collector and Pyrolyser (ACP) [2] instrument collected aerosol particles in the atmosphere of Titan and passed these to a separate instrument, the Gas Chromatograph Mass Spectrometer (GCMS) [3] for subsequent analysis. However, even in such cases, scientific instruments maintain their separate technical, scientific and organisational identities. Such collections of instruments predominantly, or even exclusively, occur with missions dedicated to the exploration of the Solar System and typically involve travelling to the target of interest and making a series of observations remotely, often in close proximity to the target, or in-situ. Examples of the former include missions such as Venus Express Mission (VEX) [4], BepiColombo [5] and Solar and Heliospheric Observatory (SOHO) [6] while the latter encompasses such missions as Rosetta [7] and Huygens [1]. Occasionally, a mission will include instruments for both in-situ and remote observations such as Giotto [8] and Solar Orbiter [9].

Alternatively, there are observatory missions. These typically involve a telescope-type payload with one or more instruments deployed to analyse the light collected in whatever region of the electromagnetic spectrum that is under investigation. A variant is to have a small collection of, often co-aligned, instruments to study the same target or region of sky. Unlike the category previously described, these missions are almost exclusively of an astrophysical nature, normally operating in those parts of the electromagnetic spectrum inaccessible from the Earth's surface and using space as a convenient location from which to collect and analyse the relevant information. Examples of such missions include INTEGRAL (Gamma-ray) [10], XMM-Newton (X-ray) [11] and Herschel (infra-red and sub-mm) [12].

Another type of mission are those such as Gaia [13] where the payload was procured directly by ESA and the scientific community is responsible for the scientific data processing – a task comparable in scope to delivering a major instrument in this case. It should be noted that instrument consortia often provide parts of the scientific ground segment of ESA's Science Directorate missions.

Whichever the mission type, instruments are usually provided by consortia of scientists from one or more countries both from the ESA Member States and beyond. Instruments vary in size and complexity from simple through to extremely large and complex costing many tens of million of euros. Instruments are usually selected competitively guided by the ESA Science Advisory Structure. This selection process can be complex, involving a whole range of factors including scientific and technical excellence, organisational and managerial feasibility and financial viability. Issues of industrial return, and international balance can also play a significant role. In the case where the mission involves collaboration with other agencies (i.e., National Aeronautics and Space Administration (NASA), Japanese Aerospace Exploration Agency (JAXA), China National Space Agency (CNSA), etc), then the national balance between the relevant agencies will also come into play. Selection of an instrument consortium implies a degree of recognition and trust in that consortium and in the consortium leader, or leaders, who will by implication be regarded as having significant scientific standing in the scientific area of the particular mission as well as the necessary scientific, technical, managerial and organisational skills required to bring the instrument to fruition. Thus, selection as a Principal Investigator (PI), as the consortium leader is usually referred to, undoubtedly conveys a strong degree of recognition within the space science community and the wider research domain.

Since the process of payload selection has been in progress for approaching 50 years in a relatively homogeneous way, this offers the opportunity for an analysis of the results of such selections from the perspective of at least the nation of the PI's host institution and other factors such as age and gender, although in the latter case, the relative paucity of numbers makes this more difficult. This is an area of investigation that has been relatively neglected to date yet has the potential to provide interesting and potentially instructive insights into the process, its outcomes and implications. [14] analysed the participation of women scientists in 10 ESA Solar System missions selected between 1980 and 2012 finding between 4% and 25% participation with several missions with no women PIs. Biases, both conscious and unconscious, may be at play in the process but that is not the underlying premise of this study. The whole process of payload selection is complex and depends on many interrelated factors. Understanding these and improving the process, if that is indeed possible, partly depends on data collection and analysis of the type that is presented here. We do not claim any definitive answers at this stage but believe that analysis of this type may help to understand better some of the factors at play and to improve the process where possible.

We note that not all the current Member States have been members since the start of ESA. The eleven founding Member States of Belgium, Denmark, France, West Germany, Ireland, Italy, the Netherlands, Spain, Sweden, Switzerland, and the United Kingdom joined in 1975. These were followed by Austria in 1984, Norway in 1985, Finland in 1994, Portugal in 2000, Greece and Luxembourg in 2004, the Czech Republic in 2008, Romania in 2011, Poland in 2012, and Estonia and Hungary in 2015. Science Programme payload selections are normally open worldwide, including countries that are not yet Member States. However, for the non-founding Member States, their first payload PI or Co-Principal Investigator (co-PI) was usually appointed some years after becoming members of ESA. Two exceptions are Finland which joined ESA in 1995 and had a payload PI on SOHO whose payload was selected in 1989 and Hungary which became a member in 2015 and had a PI on Rosetta whose payload was selected in 1993 (see Table 1).

## 2 Methodology

Data were collected on 28 ESA-led missions in the ESA Science Programme from a variety of sources including mission web sites, press releases, papers published in the scientific literature and the NASA Space Science Data Coordinated Archive (`nssdc.gsfc.nasa.gov`). The interval covered started with COS-B [15] selected in 1969 and launched in 1975 and continued to include those missions where the payload consortia had been selected by the end of 2021 including PLATO [16] JUICE [17], Comet Interceptor [18] and ARIEL [19]. A total of 422 PIs and co-PIs were identified. We note that the Exosat [20] payload was procured directly by ESA and so is not considered here. The Gaia [13] and Hipparcos payloads were also procured by ESA, but there were substantial national funding contributions to the data processing. These were comparable in scale to the provision of a major instrument and so these missions are included. Both Cluster and GEOS [21] were re-flights, with identical payloads, of earlier missions that failed to achieve their nominal missions — in the case of Cluster due to a launch failure and in the case of GEOS due to a non-nominal final orbit. For each considered mission, a list of instruments was obtained and for each instrument the following information was collected:

1. The names of the PI or PIs and where appropriate any co-PIs
2. The institutes of the PI(s) and any co-PIs
3. The countries of the PI(s) and any co-PIs institutes
4. The gender of the PI(s) and any co-PIs
5. The year of award of PhD degree of the PI(s) and any co-PIs

It should be noted that the consortium that is selected initially often evolves over time, particularly since the mission duration from initial selection to completion may be as long as ~20 years. Changes can occur in consortium composition due to a variety of reasons including retirement of the PI, changing financial, personnel or organisational environment, or other reasons. For this analysis, it was decided to use data pertaining only to the original selections. This left 336 PIs and co-PIs from the ESA-led missions listed in Table 1 of which 279 were located in the ESA Member States. This selection represents the thinking of the Advisory Structure and associated infrastructure of ESA and the Member States and thus any biases, conscious or unconscious, are more likely to be contained in those data rather than those resulting from any subsequent evolution of the relevant consortia.

The genders of PIs and any co-PIs were assigned through the personal knowledge of the authors, or the relevant project scientists, or when given on the personal web pages of the individuals concerned, or from other sources (listed below). We appreciate that gender identity is more complex than a binary issue. However, no attempt was made to assign genders other than male or female as this information is not readily available.

In order to determine the "age" or experience of PIs and any co-PIs, we used the year that they obtained their Doctor of Philosophy Degree (PhD) as a proxy. The was determined for the majority of those concerned by searching the internet, particularly sites such as the Astrophysics Data Service (ADS), LinkedIn, the Astronomy Genealogy Project (`astrogen.aas.org`), ORCID.org, IEEE Xplore (`https://ieeexplore.ieee.org/Xplore/home.jsp`), and, for French theses `https://www.theses.fr`. Some proposers were contacted and provided their PhD dates by email. The proposers for which the PhD could not be found are often retired, deceased or have left astronomy. For (co-)PIs who had not yet completed their degrees, the expected year of submission was used. For the small number of proposers who did not have a PhD and were not enrolled in a PhD programme, their dates were assumed to be arbitrarily far in the future. For some late career scientists in Italian institutes who do not have a PhD, their "age" was taken to be three years after they obtained their Laurea. We note that using the year of PhD to indicate the number of years experience neglects time spent outside of astronomy.

## 3 Payload Contributions

Table 1 shows the numbers of original PIs and co-PIs located in the ESA Member States for the ESA-led missions of the Science Programme. These are shown plotted in Figs. 2 and 3. As well as the 279 original PIs and co-PIs from institutes within the ESA Member States, there are an additional 56 PIs and co-PIs located outside the ESA Member States, mainly in the US and Japan with 26 and 15 PIs and co-PIs, respectively. Missions such as Proba 2 [22] and ExoMars [23] where the payloads were not selected by the ESA Science Advisory Structure were excluded from this analysis. All ESA Member States except for Greece, Luxembourg and Romania have PIs or co-PIs located at institutes in their countries with Estonia having its first co-PI in 2019 (Comet Interceptor) and Portugal (ARIEL) in 2020. The four largest financial contributors to the ESA Science Programme (Germany, the United Kingdom, France and Italy) dominate the provision of payloads with 62, 38, 53 and 39 PIs and co-PIs, respectively. This makes a total of 192 PIs and co-PIs which is 69% of the total located in the ESA Member States. In contrast the same four Member States contributed 63% of the Science Programme funding. The Member States with the next largest number of original PIs and co-PIs are Switzerland and Sweden with 16 and 15, respectively.

The time dependence of the selection of payload PIs and co-PIs is shown in Fig. 1. This shows the number of PIs and co-PIs that were selected for the years in which selections took place. The percentages of these PIs and co-PIs that are female is also shown. It is notable that there is generally a very low female participation rate until around 2014 with many of the early selections having no female PIs or co-PIs. From around 2014 ~25% of the selected PIs and co-PIs are female probably reflecting the population of female astronomers measured by the International Astronomical Union (IAU) (see Table 2).

**Table 1** The number of originally selected PIs and co-PIs contributing to payload or science operations from each ESA Member State. % Fin. Contrib. is the percentage financial contribution between 2000 and 2021 of Member States to the Science Programme. Class refers to Small/Smart, Medium or Large sized mission class in the ESA Science Programme. Year is the year of mission or payload selection. Ratio is the fraction of publications from a Member State divided by the financial contribution to the Science Programme from that Member State. The Weighted Ratios include factors for average co-PI contributions (assumed to be one half that of a PI), mission class (factors of 2.0, 1.0 and 0.3 were applied for Large, Medium and Small/Smart missions), and inversely to the total number of PIs and co-PIs that a mission has (including from non-ESA Member States).

| Mission | Class | Year | AT | BE | CH | CZ | DE | DK | EE | ES | FI | FR | GB | GR | HU | IE | IT | LU | NL | NO | PL | PT | RO | SE | Total ESA MS |
|---|---|---|---|---|---|---|---|---|---|---|---|---|---|---|---|---|---|---|---|---|---|---|---|---|---|
| COS-B | M | 1969 | | | | | 1 | | | | | 1 | | | | | 2 | | 1 | | | | | | 5 |
| GEOS | M | 1970 | | | 1 | 4 | 1 | | | | 3 | 1 | | | | 1 | | | | | | | | | 12 |
| Ulysses | M | 1977 | | | 1 | | 6 | | | | | 1 | 1 | | | | 1 | | | | | | | | 9 |
| Giotto | M | 1980 | | | 1 | | 4 | | | | | 2 | 2 | | | 1 | | | | | | | | | 11 |
| Hipparcos | M | 1980 | | | | | | 1 | | | | 2 | | | | | | | | | | | | 1 | 4 |
| ISO | L | 1983 | | | | | 1 | | | | | 1 | 1 | | | | | | 1 | | | | | | 4 |
| Huygens | M | 1988 | | | | | 1 | | | | | 1 | 2 | | | | 1 | | | | | | | | 5 |
| SOHO | M | 1989 | | | 1 | | 4 | | | | 1 | 3 | 2 | | | | 1 | | | 1 | | | | | 13 |
| Cluster | M | 1989 | 1 | | | | 2 | | | | | 3 | 3 | | | | | | | | | | | 1 | 10 |
| XMM-Newton | L | 1991 | | | | | 1 | | | | | | 2 | | | | 1 | | 1 | | | | | | 5 |
| Rosetta | L | 1993 | 2 | | 1 | | 11 | | | | 1 | 6 | 3 | | 2 | | 3 | | | | | | | 2 | 31 |
| INTEGRAL | M | 1993 | | | 1 | | 1 | 1 | | 1 | | 3 | | | | | 2 | | | | | | | | 9 |
| Planck | M | 1996 | | | | | | 1 | | | | 2 | | | | | 2 | | | | | | | | 5 |
| Herschel | L | 1998 | | 1 | | | 2 | | | | | 2 | 1 | | | | | | 1 | | | | | | 7 |
| SMART-1 | S | 1999 | | | 1 | | 3 | | | | 2 | | 1 | | | | 2 | | | | | | | | 9 |
| MEX | M | 1999 | | | | | 2 | | | | | 2 | | | | | 2 | | | | | | | 2 | 8 |
| Gaia | M | 2000 | | | | | | | | | | | | | | | 1 | | 1 | | | | | | 2 |
| BepiColombo | L | 2000 | 2 | | 1 | | 4 | | | | 3 | 6 | 4 | | | | 7 | | 1 | | | | | 3 | 31 |
| Solar Orbiter | M | 2000 | | 1 | 2 | 1 | 5 | | | 2 | | 4 | 5 | | | | 2 | | | 1 | | | | | 23 |
| VEX | M | 2002 | 1 | 1 | | | 3 | | | | | 3 | | | | | 2 | | | | | | | 1 | 11 |
| LISA PF | S | 2004 | | | | | 1 | | | | | | | | | | 1 | | | | | | | | 2 |
| Euclid | M | 2012 | | | | | | | | | | 1 | | | | | | | | | | | | | 1 |
| JUICE | L | 2013 | | | 1 | 1 | 3 | | | | | 2 | 2 | | | | 4 | | 1 | | 1 | | | 2 | 17 |
| CHEOPS | S | 2014 | | | 3 | | | | | | | | | | | | | | | | | | | | 3 |
| PLATO | M | 2014 | | | | | 1 | | | | | | | | | | | | | | | | | | 1 |
| SMILE | S | 2015 | 1 | | | | | | | | | 1 | 3 | | | | | | | | | | | | 5 |
| Comet Interceptor | S | 2019 | | | 2 | 1 | 1 | | 1 | 1 | 1 | 2 | 2 | | | | 2 | | | | 1 | | | 1 | 15 |
| ARIEL | M | 2020 | 1 | 1 | | 1 | 1 | 1 | 1 | 2 | | 2 | 2 | 1 | 1 | | 2 | | 1 | 1 | 1 | 1 | | 1 | 21 |
| Total | | | 8 | 4 | 16 | 4 | 62 | 5 | 2 | 6 | 8 | 53 | 37 | 3 | 2 | | 39 | | 8 | 3 | 3 | 1 | | 15 | 279 |
| % ESA Payload | | | 2.87 | 1.43 | 5.73 | 1.43 | 22.22 | 1.79 | 0.72 | 2.15 | 2.87 | 19.00 | 13.26 | | 1.08 | 0.72 | 13.98 | | 2.87 | 1.08 | 1.08 | 0.36 | | 5.38 | |
| % Fin. Contrib. | | | 2.17 | 2.72 | 3.41 | 0.92 | 21.16 | 1.74 | 0.12 | 7.22 | 1.33 | 14.94 | 15.41 | 1.45 | 0.62 | 1.03 | 11.63 | 0.20 | 4.42 | 2.16 | 2.65 | 1.17 | 0.95 | 2.58 | |
| Ratio | | | 1.32 | 0.53 | 1.68 | 1.55 | 1.05 | 1.03 | 6.03 | 0.30 | 2.15 | 1.27 | 0.86 | | 1.74 | 0.70 | 1.20 | | 0.65 | 0.50 | 0.41 | 0.31 | | 2.08 | |
| Weighted Ratio | | | 0.66 | 0.45 | 1.08 | 0.56 | 1.11 | 1.62 | 1.96 | 0.16 | 1.19 | 1.42 | 0.88 | | 0.99 | 0.48 | 1.24 | | 1.85 | 0.18 | 0.17 | 0.14 | | 1.65 | |

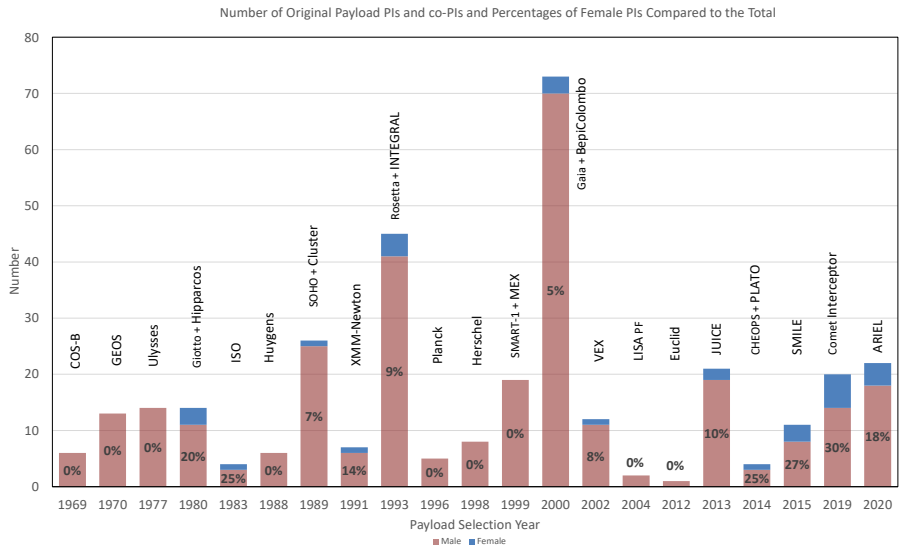

**Fig. 1** The number of payload PIs and co-PIs for the years in which selections took place. The percentages are the number of female PIs and co-PIs divided by the total number of PIs and co-PIs for each selection year. The increase in the percentage of female PIs and co-PIs from around 2014 to around ~25% of the total is apparent.

The wide range in the number of PIs and co-PIs between different missions is evident in Table 1; there are 31 from the ESA Member States on BepiColombo [5] and only one each on PLATO [16] and Euclid [24]. In general, the planetary missions tend to have more PIs and co-PIs than the observatory missions. Table 1 shows the number of PIs and co-PIs from each ESA Member State for each mission. These numbers can be compared to the average relative contributions to the ESA Science Programme funding between 2000 and 2021 which we take to be the expected relative payload contribution. The ratio of these two numbers gives an indication of whether a particular ESA Member State has provided more or less payload elements than expected. These ratios are listed in the bottom section of Table 1 and plotted against the percentage contribution in Fig. 4.

Examining Fig. 4 and Table 1 shows that of the "big four" contributors, who between them provide 63% of the Science Programme funding cluster together on the figure. It can be seen that Germany has a payload contribution comparable to its financial contribution, The United Kingdom provides less payload contribution than expected and both Italy and France provide more payload contributions than expected. The fifth biggest funding contributor is Spain which has a very low payload contribution, compared to expectation. It is more difficult to draw conclusions for some of the smaller countries due to the small number of original PIs and co-PIs involved.

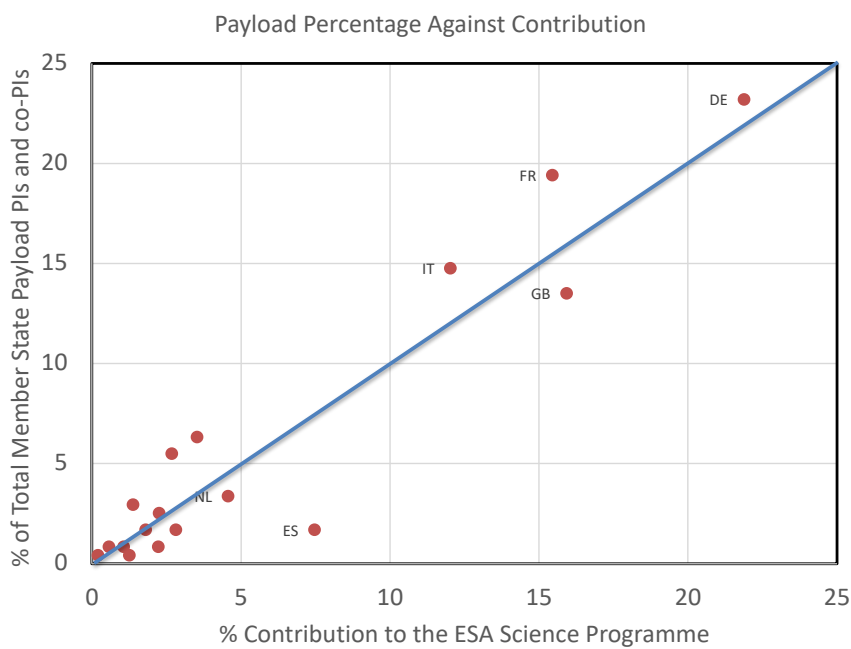

**Fig. 2** The percentages of PIs and co-PIs providing payload contributions to the ESA-led missions from ESA Member States plotted against their 2000–2021 financial contributions. Member States without PIs or co-PIs are not shown.

## 4 Payload Complexity and Size

We are aware that simply counting the number of PIs and co-PIs may bias the results towards Member States that support relatively more planetary missions compared to astronomy missions due to the larger number of payload elements on planetary missions. For that reason we have recalculated the ratios of original PI and co-PI numbers compared to the financial contributors by first applying a series of weighting factors. We assigned an overall weighting factor to each mission. This is the product of three different factors:

1. The mission class
2. The number of worldwide PIs and co-PIs for a mission
3. The relative contribution of the average co-PI compared to a PI

### 4.1 Mission Class

A complicating factor in any such comparative study is the fact that space science missions have different complexities, sizes and therefore costs. In order to investigate whether such differences could affect the analysis of the payload contributions presented above, all of the missions under consideration have been ascribed a weighting factor. The procedure that was chosen mirrors the broad system of mission classification which has been used within the ESA Science Programme in recent years. The ESA Science Programme has been mostly centred around so-called M, L & S class missions where these designations stand for Large, Medium and Smart/Small class missions. In order to reflect the differences in magnitude of such missions, we have used weighting factors of 2, 1 and 0.3 to reflect likely differences in payload costs which are broadly assumed to scale with the cost to ESA of each mission class. The designation for each mission is shown in Column 2 of Table 1 (Class).

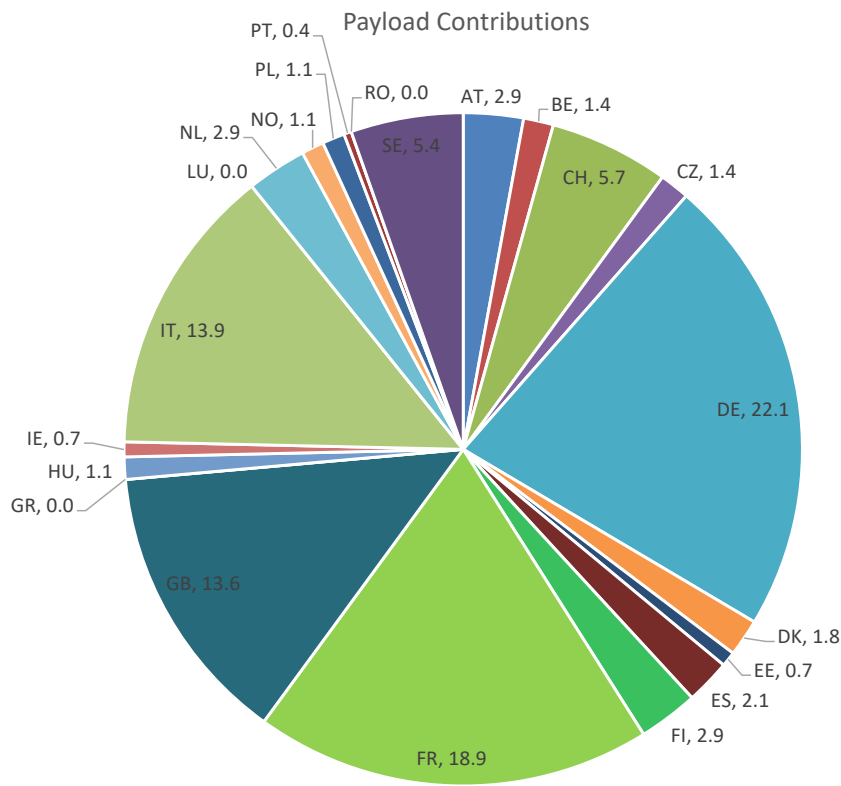

**Fig. 3** The percentages of PIs and co-PIs providing payload contributions to the ESA-led missions from ESA Member States. Member States without PIs or co-PIs are not shown.

### 4.2 Number of PIs and co-PIs

As discussed earlier, different missions can have very different numbers of instruments, varying from one or very few for observatory missions up to ten or more for typical planetary science missions. In order to account for the very different number of instruments as well as the different sizes of mission, we have divided the mission class weighting factors by the number of worldwide PIs and co-PIs (counted as half a PI, see below) before calculating the contributions of each country. The number of PIs and co-PIs for each mission is shown in the right-most column of Table 1.

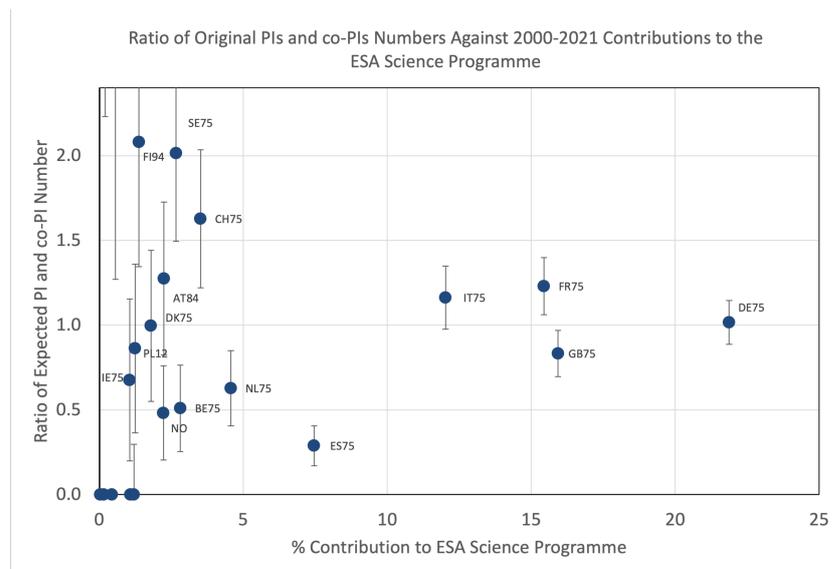

**Fig. 4** The ratio of the number of PIs and co-PIs divided by the total number from the ESA Member States normalised by the contributions of each Member State to the funding for the Science Programme plotted against the contributions. The last two digits of the year in which each country joined ESA is indicated in its label. The three Member States that have no PIs or co-PIs (Greece, Luxembourg, and Romania) are not included. The uncertainties shown are indicative only as they assume the PI and co-PI numbers follow Poisson statistics.

### 4.3 Co-PI Contributions

Most instruments have a single PI but sometimes there are a number of co-PIs. These generally reflect a funding contribution to the payload from the co-PI's nation. For the purpose of this study we have assumed that on average co-PIs contribute payload with half the value of that of a PI. We therefore included the number of PIs and 0.5 times the number of co-PIs from a particular Member State as a weighted measure of its contribution to the ESA Science Directorate missions' payloads. Figures 5 and 6 show the weighted contributions for the relevant Member States.

## 5 Effect of Weighting Factors

We then recalculated the ratios of expected payload contributions as before using the three factors described above. These are shown in the bottom line of Table 1 "Weighted Ratio" and are plotted against ESA Member State funding contributions in Fig. 7.

Comparison of Figs. 4 and 7 and the ratios given in Table 1 shows that the application of weighting factors has very little effect on the outcomes for the four large ESA Member States with the ratios increasing by 0.06, 0.15 and 0.04 for Germany, France and Italy, respectively. The ratio remained unchanged for the United Kingdom. These small changes indicate a robust process and that it is not strongly affected by payload complexity for Member States with $\gtrsim 30$ PIs and co-PIs. For the next largest contributor to the ESA Science Directorate funding (Spain) there is a small decrease in the ratio when weighting is ap-plied of 0.14 while the ratio increases strongly for the Netherlands from 0.65 to 1.85 when weighting is applied. This is likely the result of strong Dutch provision of observatory payloads with their typically smaller number of instruments. For the other ESA Member States it is difficult to draw reliable conclusions from the application of weighting factors due to the smaller number of PIs and co-PIs involved. However, comparison of Figs. 4 and 7 shows that the application of weighting factors tend to reduce the relative contributions of the smaller Member States.

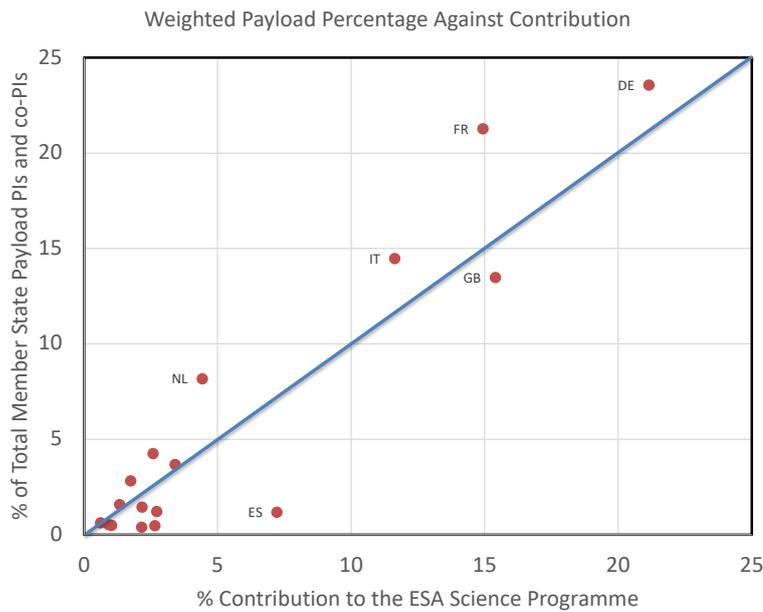

**Fig. 5** The weighted percentages of PI and co-PIs providing payload contributions to the ESA-led missions from ESA Member States plotted against their 2000–2021 financial contributions to the Programme. Member States without payload PIs or co-PIs are not shown.

## 6 Payload PI and co-PI Genders

We have examined the number of original female PIs and co-PIs compared to the total. Of a total of 336 such PIs (from Member and non-Member States) 32 were female. which is 9.5% of the total. This is much less than the fractions of female observing time PIs on the observatory missions INTEGRAL, Herschel and XMM-Newton. The clearest example is with XMM-Newton where the fraction of proposals from female PIs increases from ∼20% to ∼30% of the total over the 21 years of the mission that is being considered. This is broadly comparable with the gender distribution of the IAU membership which ranges from ∼50% female between the ages of 25–30 to ∼18% for ages 60–65 (see Table 2).

Being a PI or co-PI on an instrument can be considered to be a leadership position in the community, so these are likely to be senior people with substantial experience. To check this, we determined the average difference between the year that the payload of a mission was approved and the year in which each originally selected PIs and co-PIs received their

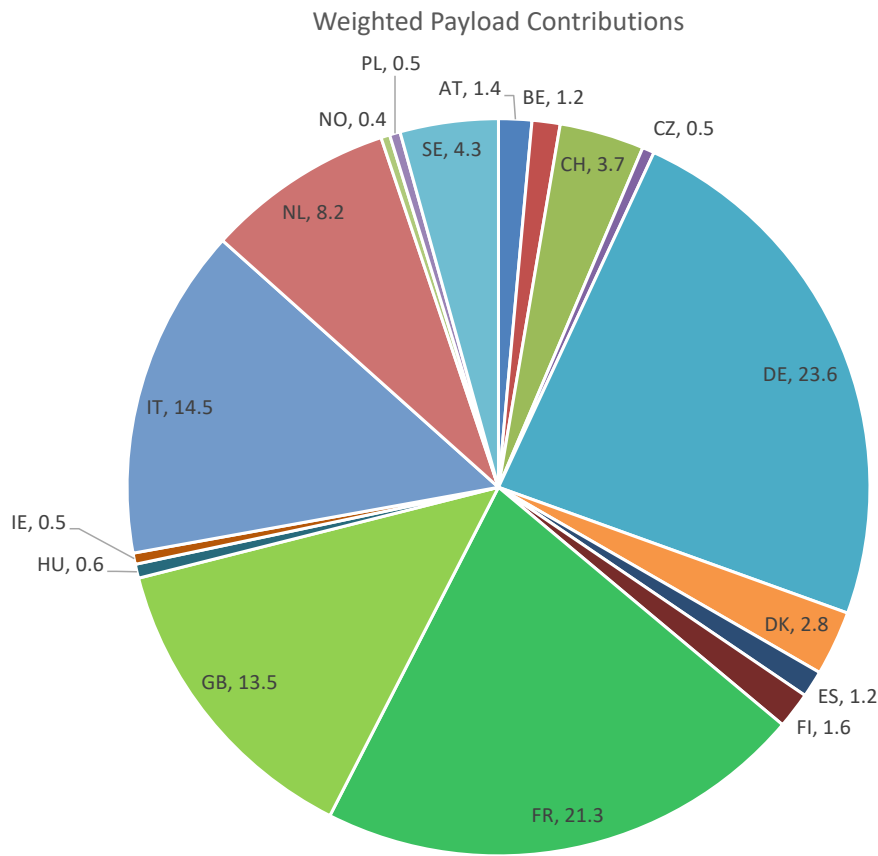

**Fig. 6** The weighted percentages of PI and co-PIs providing payload contributions to the ESA-led missions from ESA Member States. Member States without PIs or co-PIs are not shown. The method of calculating the weighting factors is described in Sect. 4.

Doctor of Philosophy Degree (PhD)s. We searched the internet, particularly sites such as the ADS, LinkedIn, the Astronomy Genealogy Project (astrogen.aas.org) and ORCID.org. For Italian astronomers without a PhD we used the year the Linea was obtained plus three years. We found that for the 84% of the 337 PIs and co-PIs where we found the year they obtained their PhD, the average difference is 15.9 years, consistent with a more "senior" population. For comparison, the average "academic ages" of XMM-Newton, INTEGRAL and Herschel observing time PIs are 10.9, 13.7 and 12.2 years, respectively (see [25], [26] and [27]). For male payload PIs the average is 15.7 years and for female PIs 17.9 years. If we assume an average age for obtaining a PhD of 25–30 years, this implies an average PI and co-PI age of 40–45 years. The current IAU membership for this age range comprises of around 26% females (see Table 2). This is similar to the percentage of female PIs and co-PIs in recent payload selections (∼25%) suggesting that the chance of selection for male and female PIs is comparable.

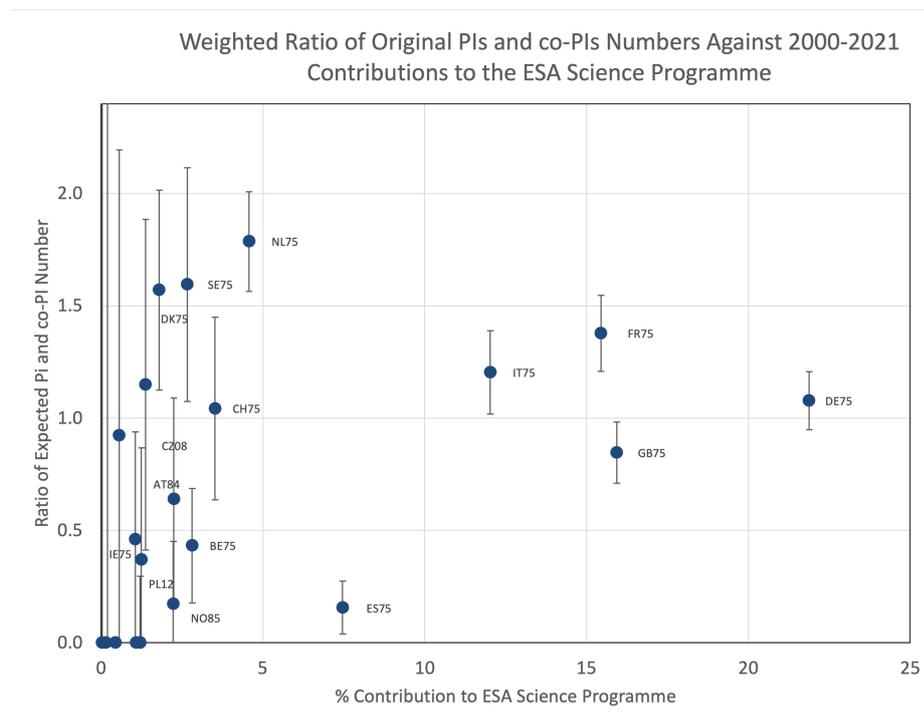

**Fig. 7** The weighted ratio of the number of PIs and co-PIs divided by the total number from the ESA Member States divided by the contributions of each Member State to the funding for the Science Programme plotted against the contributions. The year in which each country joined ESA is indicated in its label. The three Member States that have not provided payload elements (Greece, Luxembourg and Romania) are not included.

# 7 Conclusions

The number of payload PIs and co-PIs has been used as a proxy of the contributions of the different ESA Member States to the payloads of the ESA Science Directorate's missions. The four largest financial contributors to the ESA Science Programme (Germany, the United Kingdom, France and Italy) dominate the provision of payloads with in total 192 original PIs and co-PIs which is 68% of the total located in all ESA Member States. It is pleasing to see that the percentage of female PIs and co-PIs has increased in recent selections to reach ∼25%. This is comparable to the current population of "senior" female IAU members.

**Table 2** The gender distribution with age of the IAU membership of active professional astronomers. Adapted from the IAU website: www.iau.org/administration/membership/individual/distribution/.

| Age Range | Number | | | Percentage | | |
|---|---|---|---|---|---|---|
| | Female | Male | Non-Binary or Declined to Answer | Female | Male | Non-Binary or Declined to Answer |
| 25–30 | 12 | 23 | 3 | 31.6 | 60.5 | 7.9 |
| 30–35 | 189 | 296 | 20 | 37.4 | 58.6 | 4.0 |
| 35–40 | 328 | 563 | 19 | 36.0 | 61.9 | 2.1 |
| 40–45 | 415 | 873 | 23 | 31.7 | 66.6 | 1.8 |
| 45–50 | 371 | 950 | 17 | 27.7 | 71.0 | 1.3 |
| 50–55 | 307 | 1035 | 13 | 22.7 | 76.4 | 1.0 |
| 55–60 | 264 | 1149 | 8 | 18.6 | 80.9 | 0.6 |
| 60–65 | 218 | 1000 | 14 | 17.7 | 81.2 | 1.1 |
| 65–70 | 180 | 906 | 6 | 16.5 | 83.0 | 0.5 |
| 70–75 | 134 | 798 | 6 | 14.3 | 85.1 | 0.6 |
| 75–80 | 115 | 780 | 8 | 12.7 | 86.4 | 0.9 |
| 80–85 | 69 | 610 | 2 | 10.1 | 89.6 | 0.3 |
| 85–90 | 30 | 222 | 1 | 11.9 | 87.7 | 0.4 |
| 90–95 | 5 | 81 | 1 | 5.7 | 93.1 | 1.1 |
| 95–100 | 5 | 24 | 0 | 17.2 | 82.8 | 0.0 |

In order to continue studies similar to this, and to monitor any future trends, it would be invaluable if ESA and the ESA Member States would:

1. Encourage prospective instrument teams to include a breakdown of the gender distribution of their members, including the PI. This would allow much easier monitoring of the gender outcomes of the selection processes.
2. Take due account, when instrument team proposals are assessed, of the the gender of the PI and the gender distribution of the members. We expect that the number of female instrument team PIs will increase with time given the increasing fraction of early career female astronomers (see Table 2) who are now beginning to reach "senior" positions. We note that recent payload PI selections have resulted in ∼25% of female PIs, which is significantly higher than earlier selections. Nevertheless, scientific excellence should always remain the main driver for selection.
3. Establish a mentoring system for new and prospective instrument PIs which would help guide less senior scientists in this sometimes daunting task.
4. Establish an open database containing all information, in a homogeneous form, pertaining to payload teams, including institutions, academic ages, gender and any other information to enable these factors to be monitored and updated on a regular basis.

**Acknowledgements.** We thank Erik Kuulkers for helping find the PhD dates for many of the PIs and co-PIs and Peter Wenzel, Arnaud Masson, Cecil Tranquille, Steve Sembey, Michael Küppers, Philippe Escoubet, Håkan Svedhem, Matt Taylor, Johannes Benkhof and Colin Wilson for help in finding the original payload PIs and co-PIs.

## Acronym List

| | |
|---|---|
| **ACP** | Aerosol Collector and Pyrolyser |
| **ADS** | Astrophysics Data Service |
| **CNSA** | China National Space Agency |
| **co-PI** | Co-Principal Investigator |
| **ESA** | European Space Agency |
| **GCMS** | Gas Chromatograph Mass Spectrometer |
| **IAU** | International Astronomical Union |
| **JAXA** | Japanese Aerospace Exploration Agency |
| **NASA** | National Aeronautics and Space Administration |
| **PhD** | Doctor of Philosophy Degree |
| **PI** | Principal Investigator |
| **SOHO** | Solar and Heliospheric Observatory |
| **VEX** | Venus Express Mission |